\documentclass{aastex631}

\usepackage{natbib}
\usepackage[english]{babel}
\usepackage{hyperref}
\usepackage{soul}
\usepackage{xcolor}

\submitjournal{ApJ}

\shorttitle{Mrk 421 \& Mrk 501 TeV spectra with HAWC}
\shortauthors{THE HAWC COLLABORATION}

\graphicspath{{./}{figures/}}

\begin{document}

\title{Long-term spectra of the blazars Mrk 421 and Mrk 501 at TeV energies seen by HAWC}

\correspondingauthor{Sara Coutin\~no de Le\'on}
\email{sara@inaoep.mx} 
\correspondingauthor{Alberto Carrami\~nana}
\email{alberto@inaoep.mx}
\correspondingauthor{Daniel Rosa-Gonz\'alez}
\email{danrosa@inaoep.mx}
\correspondingauthor{Anna Lia Longinotti}
\email{alonginotti@astro.unam.mx}


\author[0000-0003-0197-5646]{A.~Albert}
\affiliation{Physics Division, Los Alamos National Laboratory, Los Alamos, NM, USA }

\author[0000-0001-8749-1647]{R.~Alfaro}
\affiliation{Instituto de F\'{i}sica, Universidad Nacional Autónoma de México, Ciudad de Mexico, Mexico }

\author{C.~Alvarez}
\affiliation{Universidad Autónoma de Chiapas, Tuxtla Gutiérrez, Chiapas, México}

\author{J.R.~Angeles Camacho}
\affiliation{Instituto de F\'{i}sica, Universidad Nacional Autónoma de México, Ciudad de Mexico, Mexico }

\author{J.C.~Arteaga-Velázquez}
\affiliation{Universidad Michoacana de San Nicolás de Hidalgo, Morelia, Mexico }

\author{K.P.~Arunbabu}
\affiliation{Instituto de Geof\'{i}sica, Universidad Nacional Autónoma de México, Ciudad de Mexico, Mexico }

\author[0000-0002-4020-4142]{D.~Avila Rojas}
\affiliation{Instituto de F\'{i}sica, Universidad Nacional Autónoma de México, Ciudad de Mexico, Mexico }

\author[0000-0002-2084-5049]{H.A.~Ayala Solares}
\affiliation{Department of Physics, Pennsylvania State University, University Park, PA, USA }

\author[0000-0003-0477-1614]{V.~Baghmanyan}
\affiliation{Institute of Nuclear Physics Polish Academy of Sciences, PL-31342 IFJ-PAN, Krakow, Poland }

\author[0000-0003-3207-105X]{E.~Belmont-Moreno}
\affiliation{Instituto de F\'{i}sica, Universidad Nacional Autónoma de México, Ciudad de Mexico, Mexico }

\author[0000-0002-4042-3855]{K.S.~Caballero-Mora}
\affiliation{Universidad Autónoma de Chiapas, Tuxtla Gutiérrez, Chiapas, México}

\author[0000-0003-2158-2292]{T.~Capistrán}
\affiliation{Instituto de Astronom\'{i}a, Universidad Nacional Autónoma de México, Ciudad de Mexico, Mexico }

\author[0000-0002-8553-3302]{A.~Carramiñana}
\affiliation{Instituto Nacional de Astrof\'{i}sica, Óptica y Electrónica, Puebla, Mexico }

\author[0000-0002-6144-9122]{S.~Casanova}
\affiliation{Institute of Nuclear Physics Polish Academy of Sciences, PL-31342 IFJ-PAN, Krakow, Poland }

\author[0000-0002-7607-9582]{U.~Cotti}
\affiliation{Universidad Michoacana de San Nicolás de Hidalgo, Morelia, Mexico }

\author[0000-0002-1132-871X]{J.~Cotzomi}
\affiliation{Facultad de Ciencias F\'{i}sico Matemáticas, Benemérita Universidad Autónoma de Puebla, Puebla, Mexico }

\author[0000-0002-7747-754X]{S.~Coutiño de León}
\affiliation{Instituto Nacional de Astrof\'{i}sica, Óptica y Electrónica, Puebla, Mexico }
\affiliation{Department of Physics, University of Wisconsin-Madison, Madison, WI, USA }

\author[0000-0001-9643-4134]{E.~De la Fuente}
\affiliation{Departamento de F\'{i}sica, Centro Universitario de Ciencias Exactase Ingenierias, Universidad de Guadalajara, Guadalajara, Mexico }

\author{R.~Diaz Hernandez}
\affiliation{Instituto Nacional de Astrof\'{i}sica, Óptica y Electrónica, Puebla, Mexico }

\author[0000-0002-2987-9691]{M.A.~DuVernois}
\affiliation{Department of Physics, University of Wisconsin-Madison, Madison, WI, USA }

\author[0000-0003-2169-0306]{M.~Durocher}
\affiliation{Physics Division, Los Alamos National Laboratory, Los Alamos, NM, USA }

\author[0000-0002-0087-0693]{J.C.~Díaz-Vélez}
\affiliation{Departamento de F\'{i}sica, Centro Universitario de Ciencias Exactase Ingenierias, Universidad de Guadalajara, Guadalajara, Mexico }

\author[0000-0001-5737-1820]{K.~Engel}
\affiliation{Department of Physics, University of Maryland, College Park, MD, USA }

\author[0000-0001-7074-1726]{C.~Espinoza}
\affiliation{Instituto de F\'{i}sica, Universidad Nacional Autónoma de México, Ciudad de Mexico, Mexico }

\author{K.L.~Fan}
\affiliation{Department of Physics, University of Maryland, College Park, MD, USA }

\author{M.~Fernández Alonso}
\affiliation{Department of Physics, Pennsylvania State University, University Park, PA, USA }

\author[0000-0002-0173-6453]{N.~Fraija}
\affiliation{Instituto de Astronom\'{i}a, Universidad Nacional Autónoma de México, Ciudad de Mexico, Mexico }

\author{D.~Garcia}
\affiliation{Instituto de F\'{i}sica, Universidad Nacional Autónoma de México, Ciudad de Mexico, Mexico }

\author[0000-0002-4188-5584]{J.A.~García-González}
\affiliation{Tecnologico de Monterrey, Escuela de Ingenier\'{i}a y Ciencias, Ave. Eugenio Garza Sada 2501, Monterrey, N.L., Mexico, 64849}

\author[0000-0003-1122-4168]{F.~Garfias}
\affiliation{Instituto de Astronom\'{i}a, Universidad Nacional Autónoma de México, Ciudad de Mexico, Mexico }

\author[0000-0002-5209-5641]{M.M.~González}
\affiliation{Instituto de Astronom\'{i}a, Universidad Nacional Autónoma de México, Ciudad de Mexico, Mexico }

\author[0000-0002-9790-1299]{J.A.~Goodman}
\affiliation{Department of Physics, University of Maryland, College Park, MD, USA }

\author[0000-0001-9844-2648]{J.P.~Harding}
\affiliation{Physics Division, Los Alamos National Laboratory, Los Alamos, NM, USA }

\author[0000-0002-7609-343X]{B.~Hona}
\affiliation{Department of Physics and Astronomy, University of Utah, Salt Lake City, UT, USA }

\author[0000-0002-3808-4639]{D.~Huang}
\affiliation{Department of Physics, Michigan Technological University, Houghton, MI, USA }

\author[0000-0002-5527-7141]{F.~Hueyotl-Zahuantitla}
\affiliation{Universidad Autónoma de Chiapas, Tuxtla Gutiérrez, Chiapas, México}

\author{P.~Hüntemeyer}
\affiliation{Department of Physics, Michigan Technological University, Houghton, MI, USA }

\author[0000-0001-5811-5167]{A.~Iriarte}
\affiliation{Instituto de Astronom\'{i}a, Universidad Nacional Autónoma de México, Ciudad de Mexico, Mexico }

\author[0000-0003-4467-3621]{V.~Joshi}
\affiliation{Erlangen Centre for Astroparticle Physics, Friedrich-Alexander-Universit\"at Erlangen-N\"urnberg, Erlangen, Germany}

\author[0000-0001-6336-5291]{A.~Lara}
\affiliation{Instituto de Geof\'{i}sica, Universidad Nacional Autónoma de México, Ciudad de Mexico, Mexico }

\author[0000-0002-2467-5673]{W.H.~Lee}
\affiliation{Instituto de Astronom\'{i}a, Universidad Nacional Autónoma de México, Ciudad de Mexico, Mexico }

\author{J.~Lee}
\affiliation{University of Seoul, Seoul, Rep. of Korea}

\author[0000-0001-5516-4975]{H.~León Vargas}
\affiliation{Instituto de F\'{i}sica, Universidad Nacional Autónoma de México, Ciudad de Mexico, Mexico }

\author[0000-0001-8825-3624]{A.L.~Longinotti}
\affiliation{Instituto de Astronom\'{i}a, Universidad Nacional Autónoma de México, Ciudad de Mexico, Mexico }

\author[0000-0003-2810-4867]{G.~Luis-Raya}
\affiliation{Universidad Politecnica de Pachuca, Pachuca, Hgo, Mexico }

\author[0000-0001-8088-400X]{K.~Malone}
\affiliation{Physics Division, Los Alamos National Laboratory, Los Alamos, NM, USA }

\author[0000-0001-9052-856X]{O.~Martinez}
\affiliation{Facultad de Ciencias F\'{i}sico Matemáticas, Benemérita Universidad Autónoma de Puebla, Puebla, Mexico }

\author[0000-0002-2824-3544]{J.~Martínez-Castro}
\affiliation{Centro de Investigaci\'on en Computaci\'on, Instituto Polit\'ecnico Nacional, M\'exico City, M\'exico.}

\author[0000-0002-2610-863X]{J.A.~Matthews}
\affiliation{Dept of Physics and Astronomy, University of New Mexico, Albuquerque, NM, USA }

\author[0000-0002-8390-9011]{P.~Miranda-Romagnoli}
\affiliation{Universidad Autónoma del Estado de Hidalgo, Pachuca, Mexico }

\author[0000-0002-1114-2640]{E.~Moreno}
\affiliation{Facultad de Ciencias F\'{i}sico Matemáticas, Benemérita Universidad Autónoma de Puebla, Puebla, Mexico }

\author[0000-0002-7675-4656]{M.~Mostafá}
\affiliation{Department of Physics, Pennsylvania State University, University Park, PA, USA }

\author[0000-0003-0587-4324]{A.~Nayerhoda}
\affiliation{Institute of Nuclear Physics Polish Academy of Sciences, PL-31342 IFJ-PAN, Krakow, Poland }

\author[0000-0003-1059-8731]{L.~Nellen}
\affiliation{Instituto de Ciencias Nucleares, Universidad Nacional Autónoma de Mexico, Ciudad de Mexico, Mexico }

\author[0000-0001-9428-7572]{M.~Newbold}
\affiliation{Department of Physics and Astronomy, University of Utah, Salt Lake City, UT, USA }

\author[0000-0001-7099-108X]{R.~Noriega-Papaqui}
\affiliation{Universidad Autónoma del Estado de Hidalgo, Pachuca, Mexico }

\author{A.~Peisker}
\affiliation{Department of Physics and Astronomy, Michigan State University, East Lansing, MI, USA }

\author[0000-0002-8774-8147]{Y.~Pérez Araujo}
\affiliation{Instituto de Astronom\'{i}a, Universidad Nacional Autónoma de México, Ciudad de Mexico, Mexico }

\author[0000-0001-5998-4938]{E.G.~Pérez-Pérez}
\affiliation{Universidad Politecnica de Pachuca, Pachuca, Hgo, Mexico }

\author[0000-0002-6524-9769]{C.D.~Rho}
\affiliation{University of Seoul, Seoul, Rep. of Korea}

\author[0000-0003-1327-0838]{D.~Rosa-González}
\affiliation{Instituto Nacional de Astrof\'{i}sica, Óptica y Electrónica, Puebla, Mexico }

\author{H.~Salazar}
\affiliation{Facultad de Ciencias F\'{i}sico Matemáticas, Benemérita Universidad Autónoma de Puebla, Puebla, Mexico }

\author[0000-0002-8610-8703]{F.~Salesa Greus}
\affiliation{Institute of Nuclear Physics Polish Academy of Sciences, PL-31342 IFJ-PAN, Krakow, Poland }
\affiliation{Instituto de F\'isica Corpuscular, CSIC, Universitat de Val\`encia, E-46980, Paterna, Valencia, Spain }

\author[0000-0001-6079-2722]{A.~Sandoval}
\affiliation{Instituto de F\'{i}sica, Universidad Nacional Autónoma de México, Ciudad de Mexico, Mexico }

\author[0000-0001-8644-4734]{M.~Schneider}
\affiliation{Department of Physics, University of Maryland, College Park, MD, USA }

\author{J.~Serna-Franco}
\affiliation{Instituto de F\'{i}sica, Universidad Nacional Autónoma de México, Ciudad de Mexico, Mexico }

\author[0000-0002-1012-0431]{A.J.~Smith}
\affiliation{Department of Physics, University of Maryland, College Park, MD, USA }

\author[0000-0002-1492-0380]{R.W.~Springer}
\affiliation{Department of Physics and Astronomy, University of Utah, Salt Lake City, UT, USA }

\author[0000-0001-9725-1479]{K.~Tollefson}
\affiliation{Department of Physics and Astronomy, Michigan State University, East Lansing, MI, USA }

\author[0000-0002-1689-3945]{I.~Torres}
\affiliation{Instituto Nacional de Astrof\'{i}sica, Óptica y Electrónica, Puebla, Mexico }

\author{R.~Torres-Escobedo}
\affiliation{Departamento de F\'{i}sica, Centro Universitario de Ciencias Exactase Ingenierias, Universidad de Guadalajara, Guadalajara, Mexico }

\author[0000-0002-2748-2527]{F.~Ureña-Mena}
\affiliation{Instituto Nacional de Astrof\'{i}sica, Óptica y Electrónica, Puebla, Mexico }

\author[0000-0001-6876-2800]{L.~Villaseñor}
\affiliation{Facultad de Ciencias F\'{i}sico Matemáticas, Benemérita Universidad Autónoma de Puebla, Puebla, Mexico }

\author{X.~Wang}
\affiliation{Department of Physics, Michigan Technological University, Houghton, MI, USA }

\author{T.~Weisgarber}
\affiliation{Department of Physics, University of Wisconsin-Madison, Madison, WI, USA }

\author[0000-0002-6623-0277]{E.~Willox}
\affiliation{Department of Physics, University of Maryland, College Park, MD, USA }

\author[0000-0003-0513-3841]{H.~Zhou}
\affiliation{Tsung-Dao Lee Institute, Shanghai Jiao Tong University, Shanghai, China}

\author[0000-0002-8528-9573]{C.~de León}
\affiliation{Universidad Michoacana de San Nicolás de Hidalgo, Morelia, Mexico }

\collaboration{80}{THE HAWC COLLABORATION}

\begin{abstract}

The High Altitude Water Cherenkov (HAWC) Gamma-Ray Observatory surveys the very high energy sky in the 300 GeV to $>100$ TeV energy range. HAWC has detected two blazars above $11\sigma$, Markarian 421 (Mrk 421) and Markarian 501 (Mrk 501). The observations are comprised of data taken in the period between June 2015 and July 2018, resulting in a $\sim 1038$ days of exposure. In this work we report the time-averaged spectral analysis for both sources above 0.5 TeV. Taking into account the flux attenuation due to the extragalactic background light (EBL), the intrinsic spectrum of Mrk 421 is described by a power law with an exponential energy cut-off with index $\alpha=2.26\pm(0.12)_{stat}(_{-0.2}^{+0.17})_{sys}$ and energy cut-off $E_c=5.1\pm(1.6)_{stat}(_{-2.5}^{+1.4})_{sys}$ TeV, while the intrinsic spectrum of Mrk 501 is better described by a simple power law with index $\alpha=2.61\pm(0.11)_{stat}(_{-0.07}^{+0.01})_{sys}$. The maximum energies at which the Mrk 421 and Mrk 501 signals are detected are 9 and 12 TeV, respectively. This makes these some of the highest energy detections to date for spectra averaged over years-long timescales. Since the observation of gamma radiation from blazars provides information about the physical processes that take place in their relativistic jets, it is important to study the broad-band spectral energy distribution (SED) of these objects. To this purpose, contemporaneous data in the gamma-ray band to X-ray range, and literature data in the radio to UV range, were used to build time-averaged SEDs that were modeled within a synchrotron self-Compton leptonic scenario.

\end{abstract}

\section{Introduction}\label{sec1}
Blazars are a particular class of radio-loud Active Galactic Nuclei (AGN), characterized by ultra relativistic jets escaping from a super massive black hole (SMBH) that are oriented very close to the observer’s line of sight \citep{AGNUrry}. The spectral energy distribution (SED) of blazars is characterized by two peaks: the first at low to medium energies (radio to X-ray), and the second at high energies (gamma-ray) \citep{blazarsed}. The first peak is produced by synchrotron emission from ultra-relativistic charged particles embedded in a magnetic field within the plasma jet. The high-energy peak is thought to be produced by inverse Compton (IC) scattering of low-energy photons. These low-energy photons can be assumed to come from the same charged particle population that generates the synchrotron emission (synchrotron-self Compton model, SSC) \citep{SSC} and/or from an external region, like the AGN accretion disk or the torus of dust that surrounds it (external Compton model, EC)\citep{EC}. Also, there are multiple models that use different assumptions about the nature of the charged particles that generate the electromagnetic radiation that we see from blazars. On the one hand there are the lepton models that assume that the accelerated particles are mainly electrons, and on the other hand those that assume a population mainly of hadronic particles. There are also lepto-hadronic models that consider multiple physical processes that could take place in the blazar emission zone. The relativistic jets are a key piece to understanding the nature of black holes and their environment; however, little is known about their composition, production and how gravitational energy is transported to the dissipation zone, where radiation is generated. Therefore to constrain the physical parameters of the jet, such as the size of the emission region and magnetic field, emission models have to be applied to gamma-ray observations. The very high-energy (VHE; $\gtrsim 0.1$ TeV) observations of blazars are often motivated by flaring activity, which are performed mainly by Imaging Atmospheric Cherenkov Telescopes (IACTs). However, the average activity level of blazars has been poorly characterized. Although IACTS have performed observations campaigns to characterize the properties of the average emission of blazars, the best instrumentation to get averaged spectra in long periods of time are satellites and extensive air shower (EAS) experiments since their observation times are not biased by flaring activities due to their large duty cycle.

The blazars Mrk 421 ($z=0.031$) and Mrk 501 ($z=0.034$) were first detected in the VHE range by the Whipple Observatory, Mrk 421 in 1992 above 0.5 TeV \citep{Mrk421discover} and the Mrk 501 in 1996 above 0.3 Tev \citep{Mrk501discovery}. Their gamma-ray flux has been measured extensively with IACTs during different activity periods in the 0.1-10 TeV energy range. Also both sources are continuously detected by the Large Area Telescope on board NASA's \textit{Fermi} Gamma-ray Space Telescope (\textit{Fermi}-LAT) in the 50 MeV to 1 TeV energy range \citep{4fgl}. In the literature, the quasi-contemporaneous multi-wavelength spectra have been measured mainly through observation campaigns.

Mrk 421 has been extensively studied in the VHE range. Between 2004 and 2005 MAGIC performed a total of 25.6 hours of observations during a period where Mrk 421 had low energy flux, identifying an energy cut-off and setting upper limits for energies at and above 4 TeV  \citep{Mrk421MAGIC}. With the same telescope, in 2006 the average spectra of Mrk 421 was observed during a high energy flux state for 8 nights (12.7 h in total) where the highest energy detection was reported at 3.3 TeV \citep{MAGIC2010-MRK421}. Between 2006 and 2008 observations of the source were carried out with the VERITAS telescope where different periods of activity were identified, having a total of 35.19 hours of observations, of which 9\% correspond to observations made when the source had a high activity state \citep{Mrk421VERITAS}. The ARGO-YBJ experiment used 676 days of observation to report four different averaged spectra of Mrk 421 between 2007 and 2010, where simultaneous X-ray data were used to find correlations between the different states of activity of the source; the highest energy signal was reported at 10 TeV \citep{Bartoli2011}. In 2009, the MAGIC telescope observed Mrk 421 for 27.7 hours as part of the multi-wavelength campaign organized by the \textit{Fermi} collaboration, detecting the source at 4 TeV; emission models were fit to the SED and the results differed from those obtained in previous works since they were based on observations when the source was in a state of high activity \citep{Fermi-421-SED}. Recently, using 4.5 years of data, the ARGO-YBJ experiment reported the averaged spectrum of Mrk 421 for observations performed between 2008 and 2013; the highest energy signal detected was reported at 4.5 TeV \citep{Bartoli2016}.

Mrk 501 is also a very well studied VHE source. Shortly after its detection at VHE, during a period of intense high activity in 1997, the HEGRA collaboration reported the averaged spectra from this source for 140- and 85-hour observations performed with the Cherenkov telescopes CT1 and CT2, respectively; with the highest energy detection above 10 TeV \citep{Mrk501HEGRA}. In 2005 the MAGIC telescope made observations of Mrk 501 where it reported variability in its energy flux, identifying three states of activity; the average spectrum during the period of low activity was obtained with 17.2 hours of observations, the intermediate with 11 hours and the high with 1.52 hours, with the highest energy detection at approximately 4.5 TeV \citep{Mrk501MAGIC}. Between 2005 and 2006, the TACTIC telescope observed Mrk 501 for a total of 112.5 hours, reporting an average spectrum in the energy range of 400 GeV to 6 TeV \citep{Mrk501TACTIC}. A multi-wavelength campaign to study the low activity state of Mrk 501 was carried out in 2005 where the MAGIC telescope observed the source for 9.1 hours detecting it at up to 2 TeV; the obtained SED, using \textit{Fermi} and \textit{Suzaku} data in X-rays, was fitted with an SSC model and suggested that the high activity states could be due to variations in the electron population \citep{Mrk501MAGIC-lowstate}. A few years later, the \textit{Fermi} collaboration organized in 2009 a multi-wavelength campaign to study the Mrk 501 SED, which was built with data from radio to gamma rays; the MAGIC telescope participated with 16.2 hours of observation and the VERITAS array telescopes with 9.7 hours, of which 2.4 hours were during a high activity state of the source; a SSC model was fit to the data showing that relativistic proton shock waves are related to the bulk of the energy dissipation within the source emission zone \citep{Fermi-501-SED}. 

The average spectrum of Mrk 501 was also reported by the ARGO-YBJ experiment using 1179.6 days of data taken between 2008 and 2012. This was done for different periods of activity; it was deduced that there was a correlation between the X-ray and gamma-ray fluxes, as well as a hardening of the spectra when the flux increased, thus favoring the SSC model to explain the physical processes of this source \citep{Bartoli2012}.

We remark that most of the highest gamma-ray photons from Mrk 421 and Mrk 501 were detected in periods of high activities, but the maximum gamma-ray energy in long-term averaged spectra has never exceeded the 5 TeV threshold. Also, from the works cited above we can summarize a range of physical parameters that describe the jet of each source such as the emission zone which is assumed to be spherical with a radius $R$ that moves at a relativistic speed $v=\beta c$ throughout the jet with a bulk Lorentz factor $\Gamma$, so the observed radiation is amplified by a Doppler factor $\delta=(\Gamma(1-\beta\cos\theta))^{-1}$, where $\theta$ is the jet inclination angle. For Mrk 421 the Doppler factor value, which accounts for the relativistic effects and depends on the speed of the emission zone and pitch angle of the jet, ranges from $\delta=15-50$, the magnetic field varies between $B=39-200$ mG, and the size of the emission zone ranges from $R=(0.25-5.2)\times 10^{16}$ cm. For Mrk 501 the value for the Doppler factor varies between $\delta=12-25$, the range of the magnetic field goes from $B=15-310$ mG, and the size of the emission zone varies from $R=(0.1-13)\times 10^{16}$ cm. It should be mentioned that the values for the highest $\delta$ and $B$ and the lowest $R$ values, for both sources, correspond to models fitted to observations carried out in short periods of time, mainly by IACTs.

Both Mrk 421 and Mrk 501 have been detected by the High Altitude Water Cherenkov (HAWC) Gamma-Ray Observatory above 300 GeV. Each source is observed in average during 6.2 hours per day within the field of view of the observatory. The HAWC collaboration first published the detection of these sources in \cite{hawc-mrks} which reported on the monitoring carried out over a period of 513 days of observation. The  spectral analysis did not consider the EBL attenuation and was performed taking into account only the size of the air shower, which is weakly related to the energy of the primary gamma rays. In the second HAWC catalog of VHE sources \citep{hawc-catalogue}, which comprised 507 days of observations, the detection of both sources was also reported. The energy spectrum of Mrk 421 and Mrk 501 was recently reported using a preliminary implementation of the energy estimators developed by the HAWC collaboration, according to which the spectrum of each source was divided into energy bins for the first time \citep{hawc-mrks3}. That analysis comprised a period of 837 days and was carried out by using a different framework than the one used in this work (see Section \ref{sec:fitting}).

In this work we report the observations of Mrk 421 and Mrk 501 during $\sim1038$ days of exposure and the spectral analysis above 0.5 TeV with HAWC and we include the systematic uncertainties calculation. We built a contemporaneous SED with \textit{Fermi}-LAT data with the aim of modelling it with a SSC model in order to improve the constraints on the physical parameters of the blazars jet. These results can be used to study the characteristics of secondary gamma rays that are produced by inverse Compton scattering between the cosmic microwave background and the electron-positron pairs (caused by the interaction between primary gamma rays and the extragalactic background light), and thus, to restrict/constrain the properties of intergalactic magnetic field \citep{2014ApJ...796...18A}.

The paper is organized as follows, in Section \ref{sec2} the HAWC observatory is described along with the method to measure the energy flux, in Section \ref{sec3} the energy spectra of each source is presented, in Section \ref{sec4} a comparison with previous results is made, in Section \ref{sec5} the results are used to build a multi-wavelength SED to test a one-zone leptonic blazar emission model, and finally a summary and future work are presented in Section \ref{sec6}.
\section{The HAWC Gamma-Ray Observatory}\label{sec2}
HAWC is located at latitude $+19^\circ$N and at an altitude of 4100 meters in Puebla, Mexico. It consists of 300 water Cherenkov detectors (WCD), of 7.3 m diameter and 4.5 m height spread over an area larger than 22,000 m$^2$. Each WCD is filled with 180,000 liters of water and instrumented with four photo-multiplier tubes (PMTs) that measure the arrival time and direction of cosmic and gamma-ray primaries mostly above 300 GeV, within its $\sim2$ sr field of view.

Data are divided in 9 analysis bins ($f_{hit}$) according to the fraction of PMTs that are triggered in each shower event \citep {HawcCrab}. Recently developed energy estimators are used to divide the these $f_{hit}$ bins into 12 quarter-decade energy bins covering the 0.316-316 TeV range. In this paper we use the ground parameter method presented in \cite{crab-gp}, which uses the measured charge 40 meters from the air shower axis, along with the zenith angle of the air shower, to estimate the primary gamma-ray energy. The binning scheme is identical to \cite{crab-gp}. The data used for this analysis go from June 2015 to July 2018.

\subsection{Fitting technique}\label{sec:fitting}

\begin{figure}
\gridline{\fig{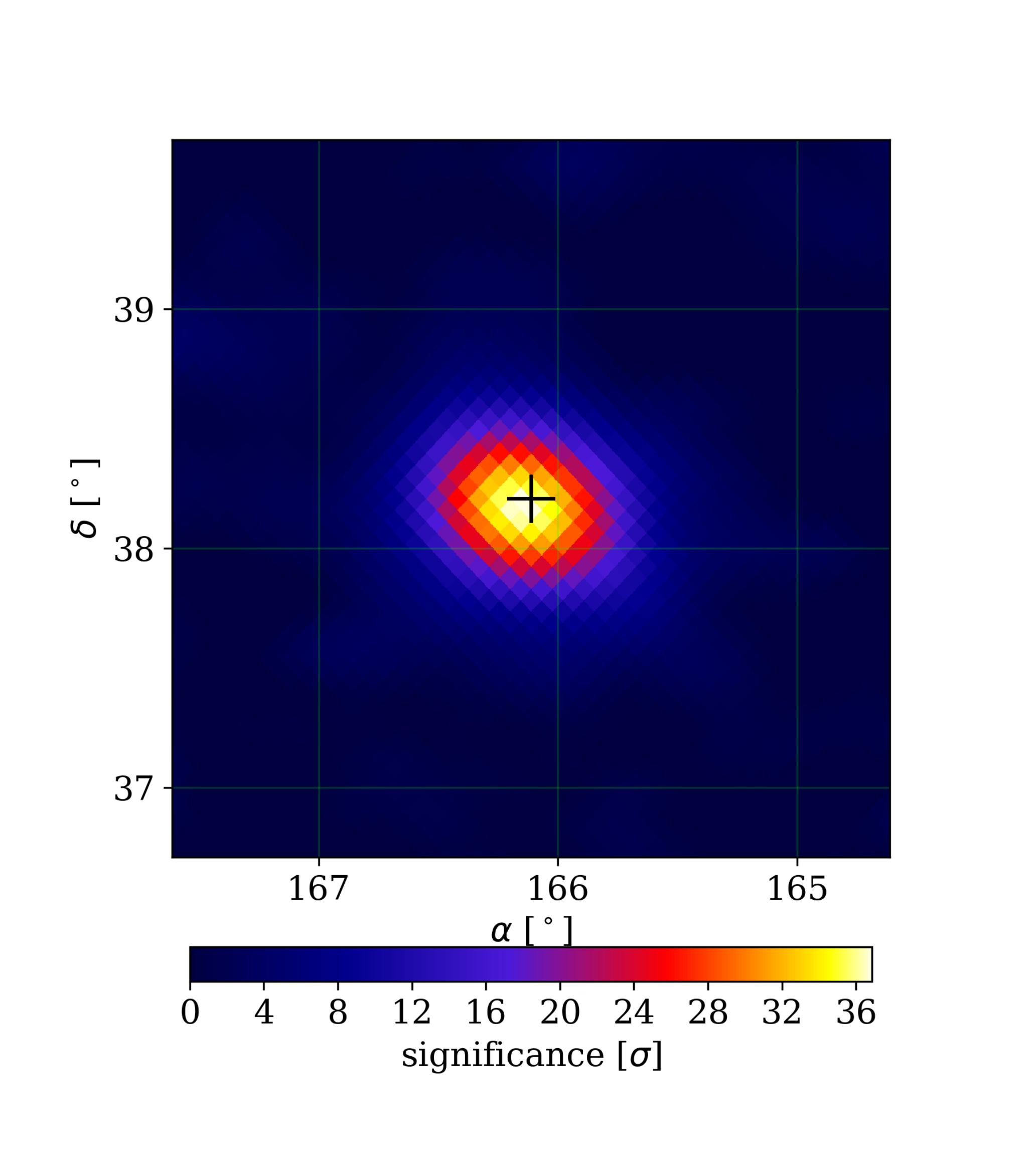}{0.49\textwidth}{Mrk 421}
          \fig{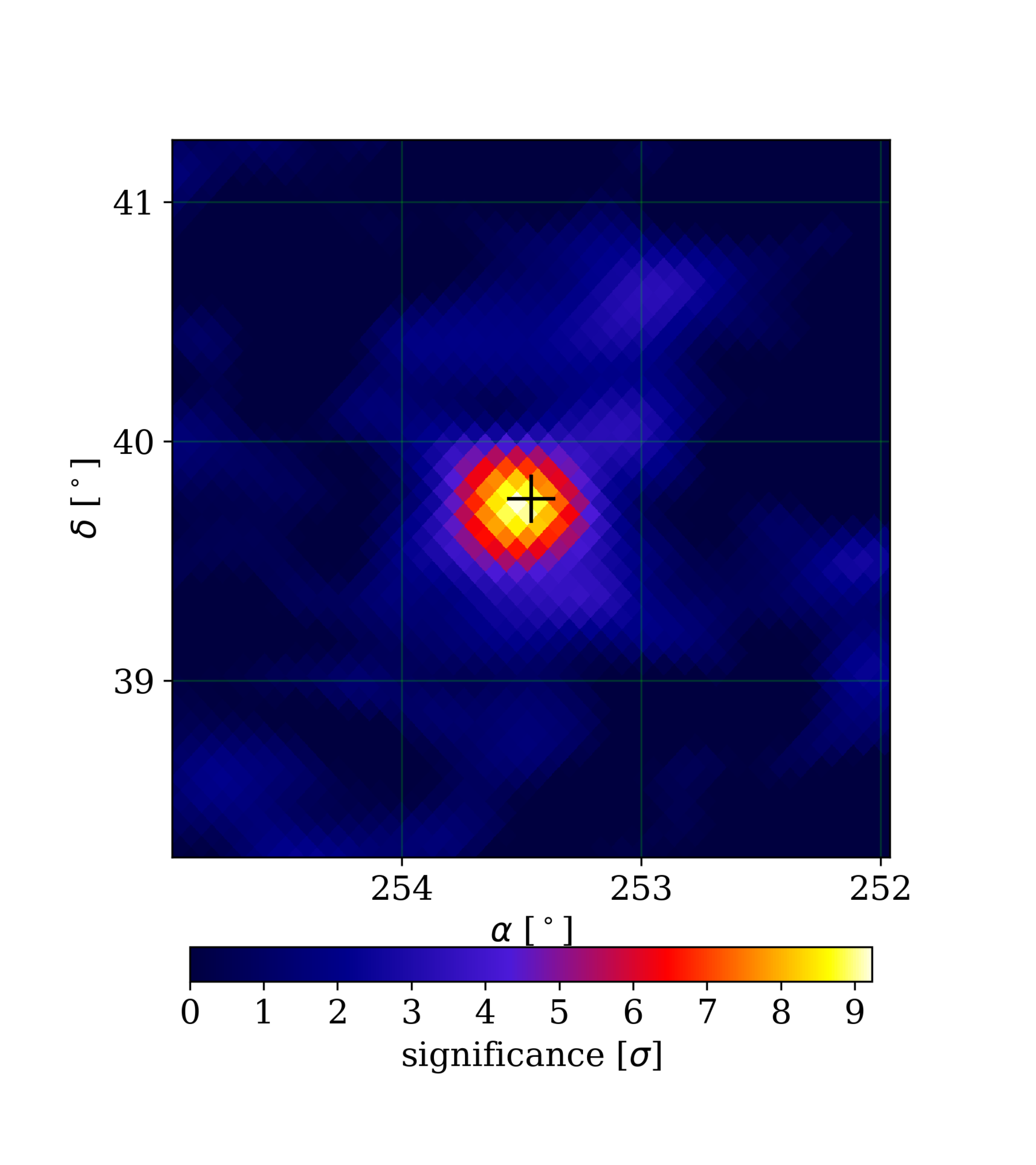}{0.49\textwidth}{Mrk 501}
          }
\caption{Significance maps of Mrk 421 (left panel) and Mrk 501 (right panel) for 1038 day of exposure corresponding to energies above 0.5 TeV, obtained with HAWC data. The cross indicates the coordinates of the source, for Mrk 421 at RA$=166.11^\circ$ and DEC$=38.2^\circ$ \citep{position421}; and for Mrk 501 at RA$=253.47^\circ$ and DEC$=39.7^\circ$ \citep{position501} equatorial J2000.0. \label{fig:sig-maps}}
\end{figure}

A forward-folding method is performed to fit the spectral shape of the sources using a maximum-likelihood technique, maximizing the test statistics ($TS$) so the input parameters have the highest likelihood of providing a good description of the observed data. Assuming a point source model, the $TS$ is defined as follows:
\begin{equation}
TS \equiv 2 \ln\frac{\mathcal{L}(H_1)}{\mathcal{L}(H_0)},
\end{equation}
where $\mathcal{L}$ is the likelihood function, $H_0$ is the background hypothesis, $H_1$ is the signal plus background hypothesis, which depends on the spectral parameters assumed to describe the sources. The significance maps (Figure \ref{fig:sig-maps}) are obtained as explained in \cite{hawc-catalogue} at the sources position, maximizing the $TS$ value in each pixel of a $N_{size}=1024$ HEALPix grid \citep{healpix}. The likelihood calculation is performed as in \cite{liff-icrc2015} using the multi-mission maximum likelihood (3ML) framework \citep{3ml} along with the HAWC accelerated likelihood (HAL)\footnote{\url{https://github.com/threeML/hawc_hal}} plugin.

Since the sources are of extragalactic origin, the attenuation due to the extragalactic background light (EBL) is taken into account. The input spectral model is assumed to be the intrinsic one and it is then attenuated using an EBL model. The resulting spectrum is then convolved with the detector response to be compared with the observed counts. This way, the output parameters correspond to the intrinsic ones. The EBL model used to perform the fits in this work is from \cite{Gil2012} .

The spectral shapes tested were a simple power law (PL, equation \ref{PL}) and a power law with an exponential energy cut-off (PL+CO, equation \ref{PL+CO}).

\begin{equation}\label{PL}
\frac{dN}{dE} = N_0 \left(\frac{E}{E_0}\right)^{-\alpha}\times \exp\left[ -\tau(E,z) \right],
\end{equation}

\begin{equation}\label{PL+CO}
\frac{dN}{dE} = N_0 \left(\frac{E}{E_0}\right)^{-\alpha}\times\exp\left(\frac{-E}{E_c}\right)\times \exp\left[-\tau(E,z) \right],
\end{equation}
where $N_0$ is the normalization flux $[\mbox{TeV}^{-1}\mbox{cm}^{-2}\mbox{s}^{-1}]$, $E_0$ is the pivot energy fixed at 1 TeV, $\alpha$ is the spectral index, $E_c$ is the energy cut-off $[\mbox{TeV}]$, and $\tau$ is the opacity value given by EBL model, and which is an increasing function of $E$ and the source redshift, $z$.

Depending on the $TS$ values in the global fit using all the available energy bins, a preferred spectral shape is chosen. The flux points are estimated as in \cite{crab-gp}, by fitting $N_0$ in each energy bin with $\alpha$ and $E_c$ fixed using the resulting values from the global fit. If a fit from an individual energy bin has a $TS<4$, an upper limit at a 95\% confidence interval is set.

\subsection{Energy range}
To determine the maximum energy at which a source is detected, the spectral model that best describes the source (nominal case) is multiplied by a step function to simulate an abrupt energy cut-off. This upper energy cut-off is set as an additional free parameter in the fit so as to provide a lower limit on the maximum detected energy when the log likelihood decreases by $2\sigma$  from the nominal case. This method has been previously used in \cite{hawc-er1}.

\subsection{Systematic uncertainties}

The size of the uncertainties depends on the detection significance of each source and spectral models chosen to describe the sources, since weaker sources naturally will show larger uncertainties and because spectral models also assign different weights in each energy bin. It is important to mention that even for sources detected at high significance, the number of free parameters affects the size of systematic uncertainties. That is, even if a source is significantly better detected than another, it will have larger systematic errors if it is fitted with a spectral model with a larger number of free parameters. The systematic uncertainties taken into account for this work were calculated as in \cite{crab-gp}, using the same simulations but with the declination of the sources of interest, we also added a 10\% in quadrature to account for additional sources of systematic errors.
\section{HAWC results}\label{sec3}
\subsection{Mrk 421}\label{subsec:421-res}

\begin{figure}[ht!]
\plotone{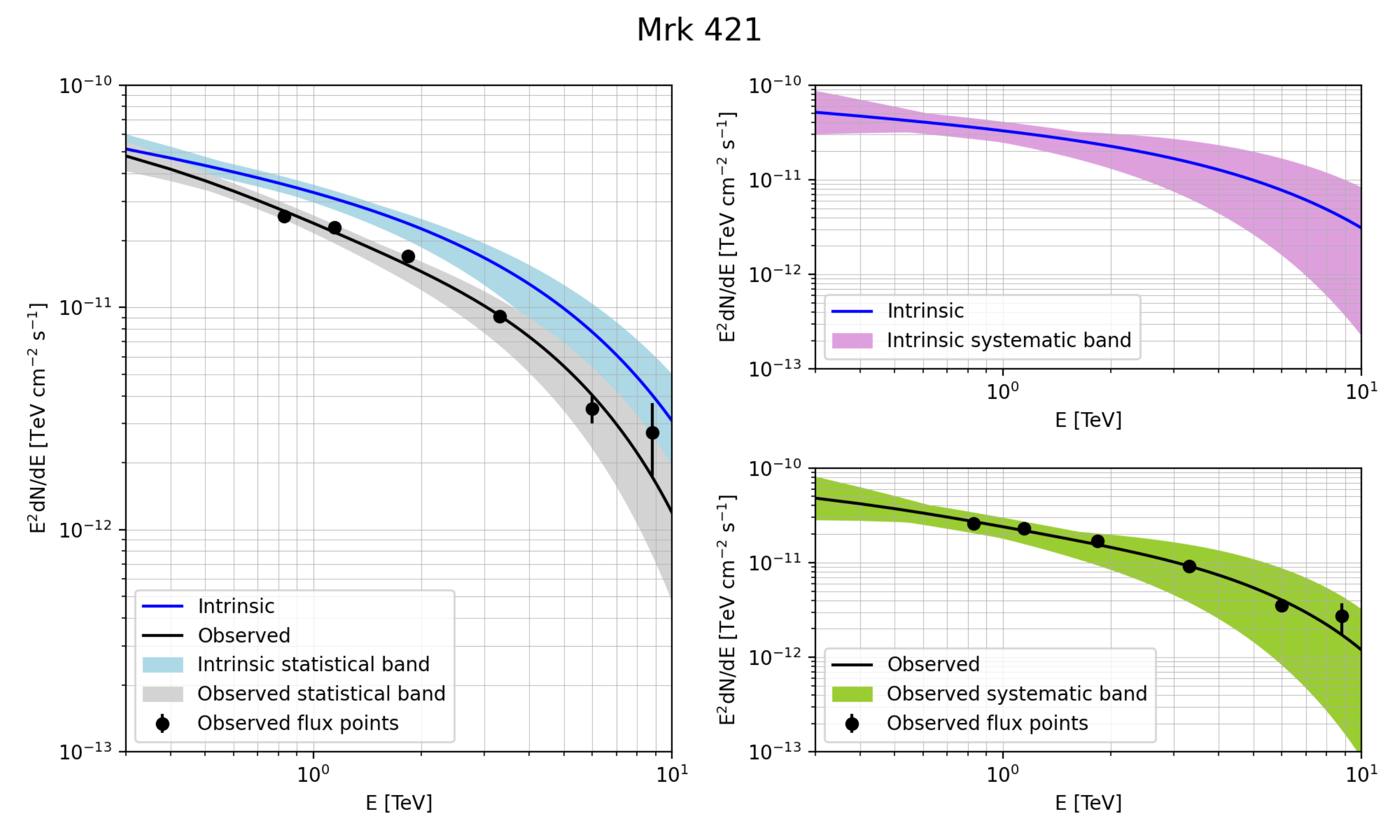}
\caption{Energy spectrum of Mrk 421. Left panel: the intrinsic spectrum is represented with the blue line along with its statistical uncertainty (blue band), the observed spectrum is represented with a black line and its statistical uncertainty (gray band) along with the observed flux points (black circles). Right panel: Intrinsic (up) and observed (bottom) spectra with their corresponding systematic band, calculated as in \cite{crab-gp}.\label{fig:spectrum-421}}
\end{figure}

All best fit parameters are quoted with their respective statistical and systematic errors. Above 0.5 TeV, the intrinsic spectra of Mrk 421 is better described by a PL+CO with $N_0=[4.0 \pm (0.3)_{stat}(_{-0.2}^{+0.9})_{sys}]\times10^{-11}\;\mbox{TeV}^{-1}\mbox{cm}^{-2}\mbox{s}^{-1}$, $\alpha=2.26 \pm (0.12)_{stat}(_{-0.39}^{+0.17})_{sys}$ and $E_c=5.10 \pm (1.60)_{stat}(_{-2.5}^{+1.4})_{sys}$ TeV (Table \ref{tab:parameters}). The intrinsic and observed differential energy spectra are shown in Figure \ref{fig:spectrum-421}. After integrating equation \ref{PL+CO} above 0.5 TeV, the observed integrated photon flux is $N_{obs}(>0.5\; \mbox{TeV})=\left[4.3\pm(0.5)_{stat}(_{-0.6}^{+0.1})_{sys}\right]\times 10^{-11}\;\mbox{ph cm}^{-2}\mbox{s}^{-1}$, and the intrinsic energy flux is $f_E(>0.5\;\mbox{TeV})=(25.7 \pm (4.2)_{stat}(^{+3.5}_{-2.6})_{sys})\times 10
^{-12}\;\mbox{erg cm}^{-2}\mbox{s}^{-1}$. The maximum energy at which the source is detected is 9 TeV at a $2\sigma$ level.

\subsection{Mrk 501}\label{subsec:501-res}

\begin{figure}[ht!]
\plotone{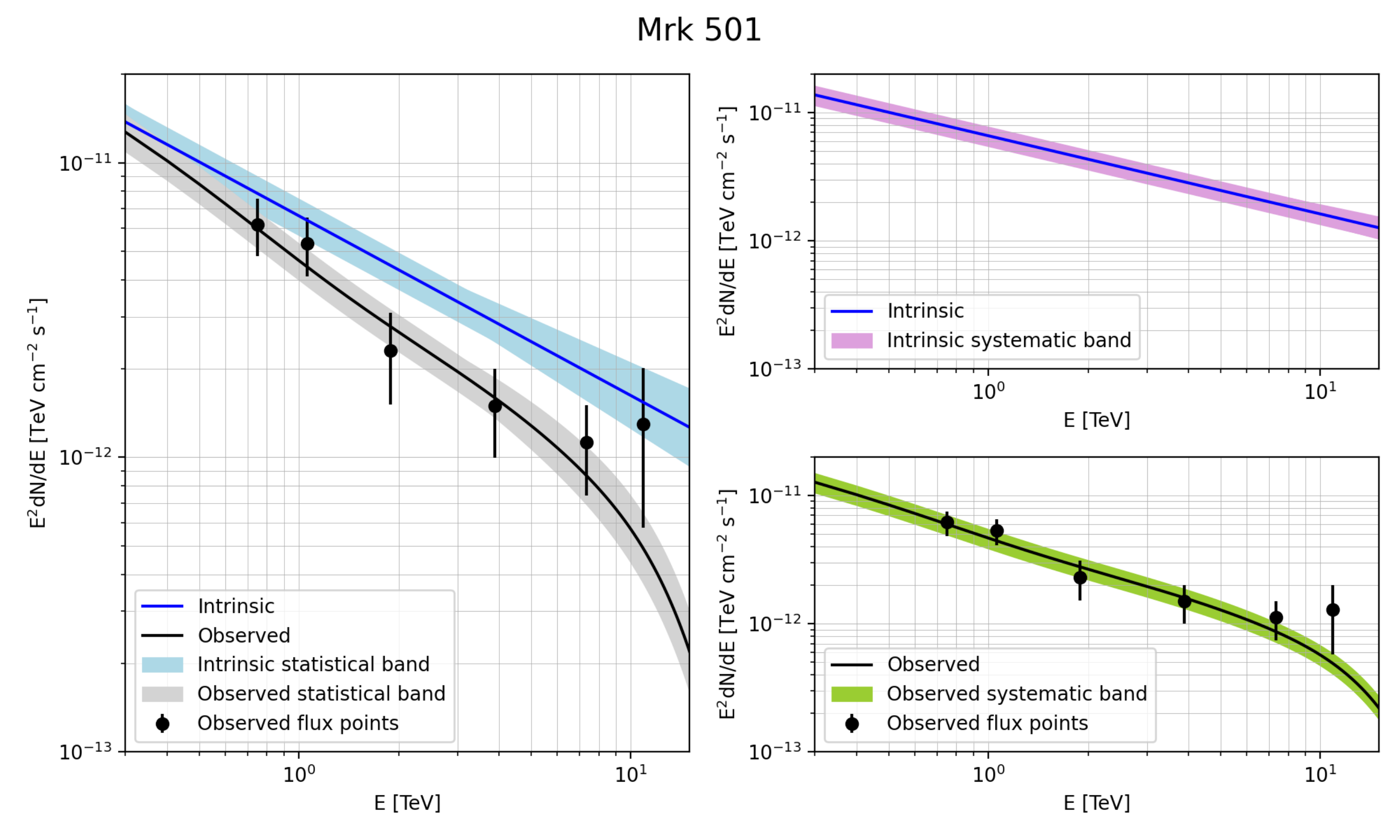}
\caption{Energy spectrum of Mrk 501. Left panel: the intrinsic spectrum is represented with the blue line and its statistical uncertainty (blue band), the observed spectrum is represented with a black line and its statistical uncertainty (gray band) along with the observed flux points (black circles). Right panel: Intrinsic (up) and observed (bottom) spectra with their corresponding systematic band, calculated as in \cite{crab-gp}.\label{fig:spectrum-501}}
\end{figure}

For Mrk 501, the best spectral model that describes de data above 0.5 TeV is a single PL with $N_0=[6.6\pm (0.9)_{stat}(_{-0.6}^{+0.9})_{sys}] \times10^{-12}\;\mbox{TeV}^{-1}\mbox{cm}^{-2}\mbox{s}^{-1}$ and $\alpha=2.61 \pm (0.11)_{stat}(_{-0.07}^{+0.01})_{sys}$. The maximum energy at which Mrk 501 is detected is 12 TeV at a $2\sigma$ level (Table \ref{tab:parameters}). Integrating equation \ref{PL} above 0.5 TeV, the observed integrated photon flux is $N_{obs}(>0.5\; \mbox{TeV})=\left[9.1\pm (1.2)_{stat}(^{+1.3}_{-0.8})_{sys}\right]\times 10^{-12}\;\mbox{ph cm}^{-2}\mbox{s}^{-1}$, and the intrinsic energy flux is $f_E(>0.5\;\mbox{TeV})=\left[25.7 \pm (4.2)_{stat}(^{+3.5}_{-2.6})_{sys}\right]\times 10
^{-12}\;\mbox{erg cm}^{-2}\mbox{s}^{-1}$. The intrinsic and observed spectra are shown in Figure \ref{fig:spectrum-501}.
 
\begin{table}[ht]
    \centering
    \caption{Best fit spectral parameters for Mrk 421 and Mrk 501 following the method described in section \ref{sec:fitting}. The photon $E_{\mbox{max}}$ value corresponds to the maximum energy at which the signal is detected at a $2\sigma$ level.}\label{tab:parameters}
    \begin{tabular}{lcc}
    \hline \hline
    & Mrk 421 & Mrk 501\\
    \hline
    $\sqrt{TS}$ & 48 & 12 \\
    $N_0\; [\mbox{TeV}^{-1}\mbox{cm}^{-2}\mbox{s}^{-1}]$ & $[4.0 \pm (0.3)_{stat}(_{-0.2}^{+0.9})_{sys}] \times10^{-11}$ & $[6.6\pm (0.9)_{stat}(_{-0.6}^{+0.9})_{sys}] \times10^{-12}$\\
    $\alpha$ & $2.26 \pm (0.12)_{stat}(_{-0.39}^{+0.17})_{sys}$ & $2.61 \pm (0.11)_{stat}(_{-0.07}^{+0.01})_{sys}$\\
    $E_c$ [TeV] & $5.1 \pm (1.6)_{stat}(_{-2.5}^{+1.4})_{sys}$ & $\infty$ \\
    
    $E_{\mbox{max}}$ [TeV] & 9 & 12 \\
    \hline \hline
    \end{tabular}
\end{table}
\section{Comparison with previous results}\label{sec4}
\subsection{Mrk 421}\label{subsec:421-comp}
\begin{figure}
\gridline{\fig{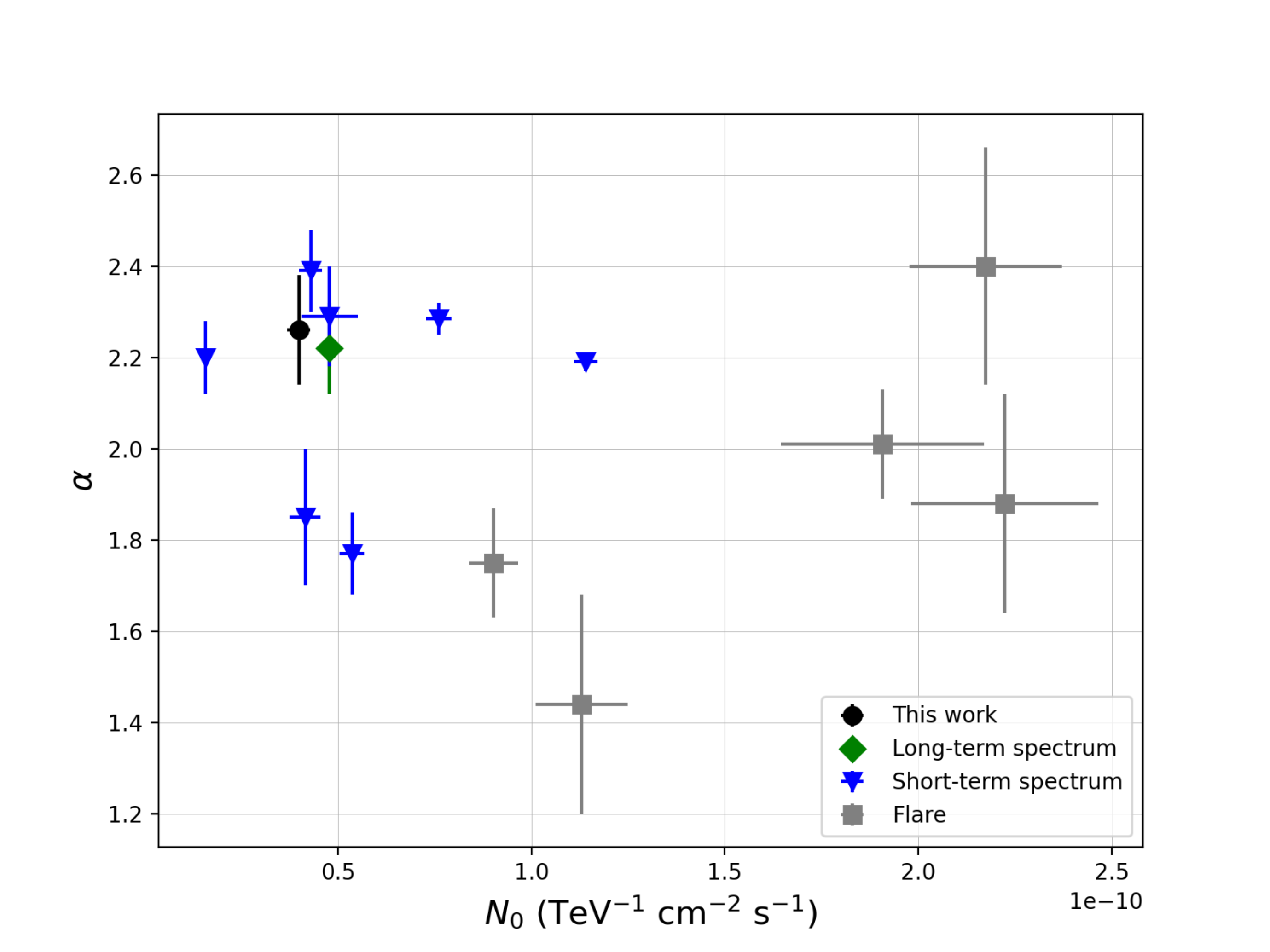}{0.49\textwidth}{Index vs. Normalization flux}
          \fig{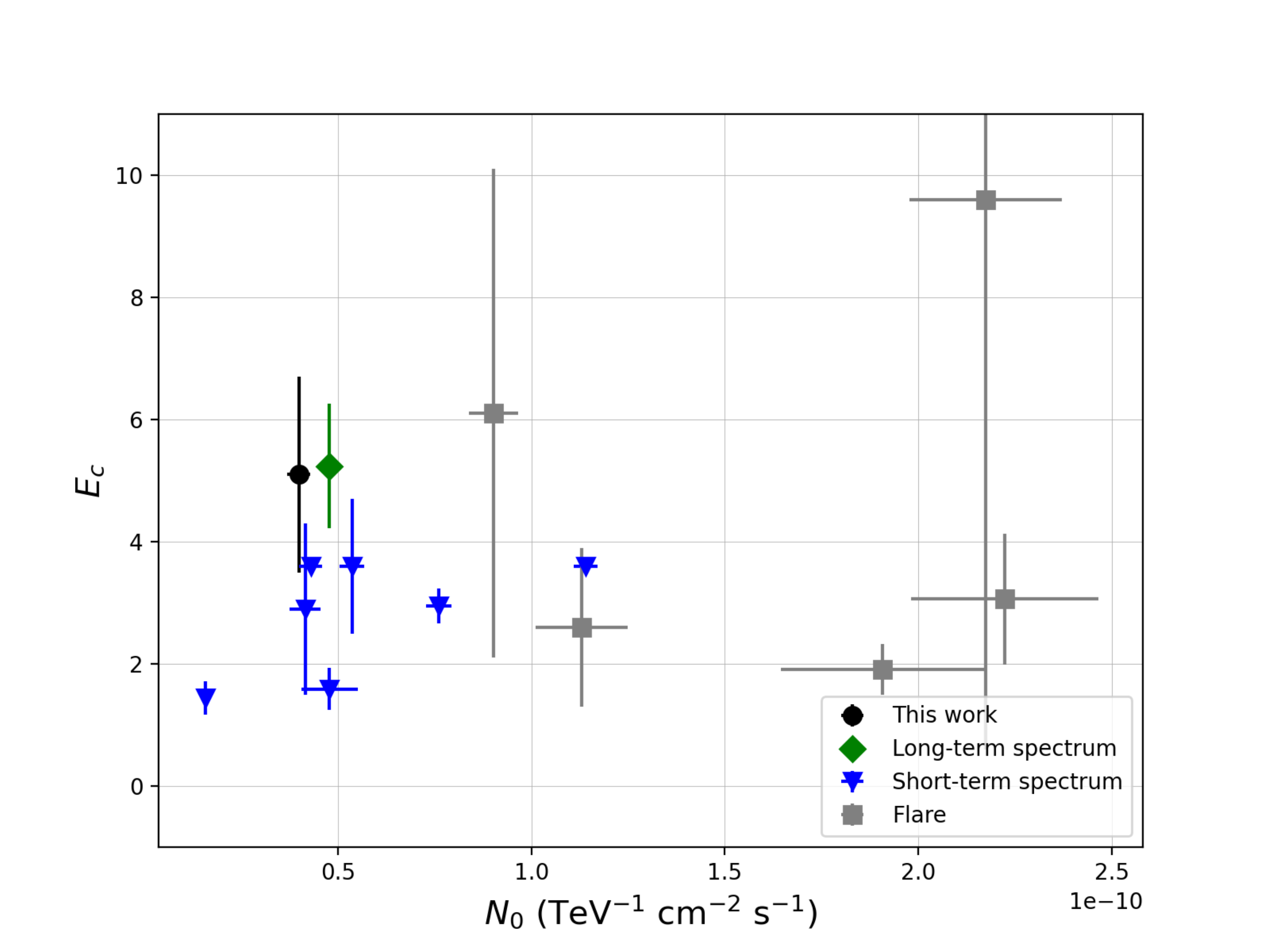}{0.49\textwidth}{Energy cut-off vs. Normalization flux}
          }
\caption{Mrk 421. Spectral index vs. normalization flux (left panel) and energy cut-off vs normalization flux (right panel) for intrinsic spectra reported values in the literature. The black circle correspond to the results in this work, the green diamonds are the previous HAWC long-term spectra reported in \cite{hawc-mrks3}, the blue downward triangles are the results reported for short-term spectra ($<1$ month) \citep{Mrk421MAGIC, MAGIC2010-MRK421, Mrk421VERITAS} and the grey squares are the reported values from observations when the source presented a high activity state \citep{MAGIC2010-MRK421, Mrk421VERITAS}.\label{fig:index-flux-421}}
\end{figure}

\begin{figure}[ht!]
\plotone{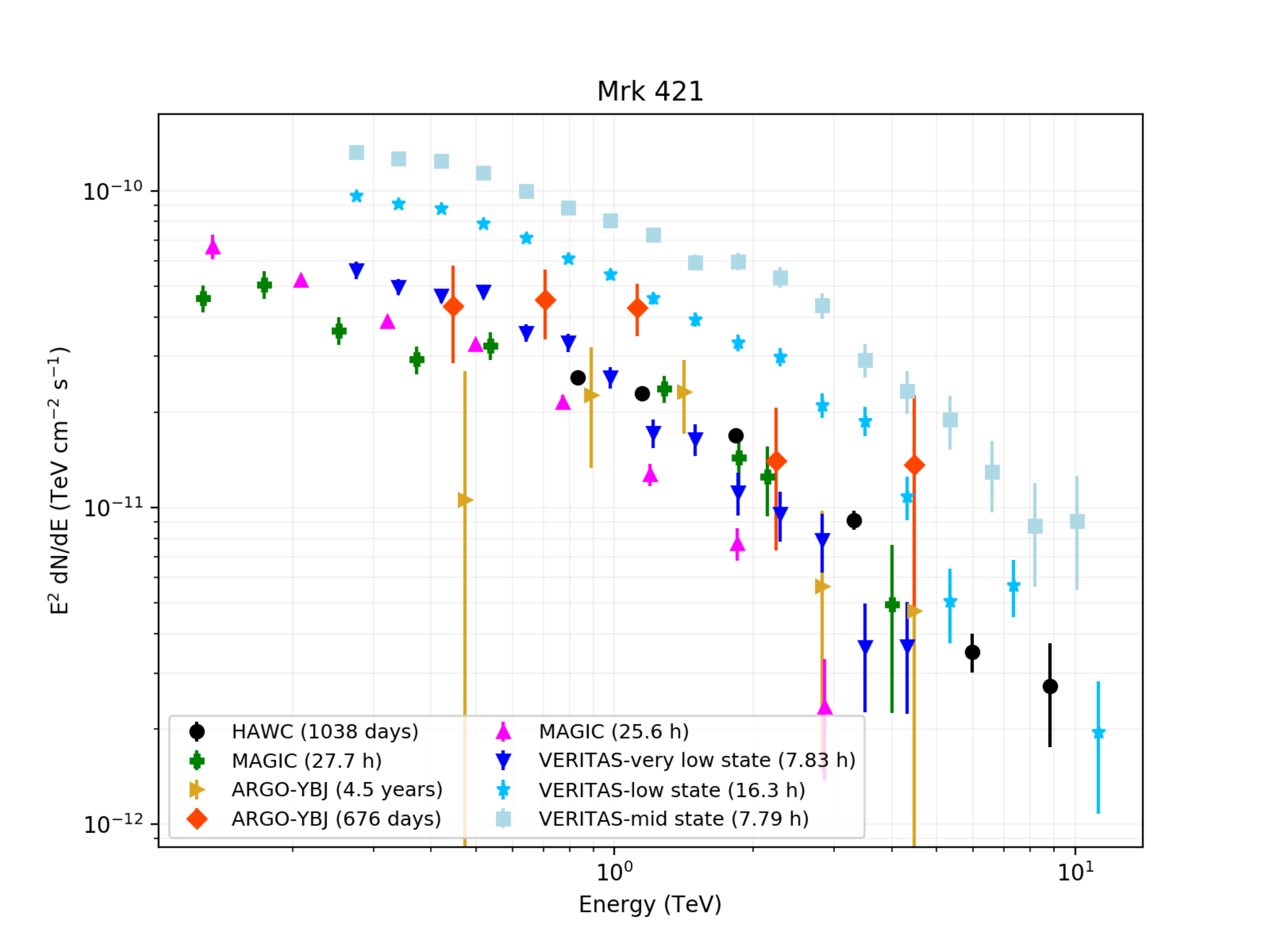}
\caption{VHE spectrum of Mrk 421. Black circles correspond to data from HAWC (1038 days), the green crosses are the MAGIC data for the 4.5-month observation campaign in 2009 \citep{Fermi-421-SED}. The right yellow triangles are the ARGOYBJ for a 4.5-year period between 2008 and 2012 \citep{Bartoli2016}. The orange diamonds are the 676 days of observations with the ARGO-YBJ experiment between 2007 and 2010 \citep{Bartoli2011}. The pink up triangles are from the 25.6-hour observations performed with MAGIC between 2004 and 2005 \citep{Mrk421MAGIC}, the blue down triangles, cian stars and light blue squares correspond to observations performed by VERITAS telescopes for a very-low (7.83 h), low (16.3 h) and mid-state (7.79 h), respectively \citep{Mrk421VERITAS}.\label{fig:421-all}}
\end{figure}

The value of the highest energy flux point in the HAWC spectrum for Mrk 421 is at 8.8 TeV. This value is one of the highest energy detections for a long-term time-averaged spectrum reported to date. The intrinsic spectrum of Mrk 421 is in good agreement within the statistical and systematic errors with the spectra previously reported by HAWC \citep{hawc-mrks3}, and with the averaged spectra reported in \cite{Mrk421MAGIC} and \cite{Bartoli2011}. As mentioned before, most of the IACT observations are biased to high-state activity of the source, showing not only a higher flux but a harder spectrum, following a “brighter-harder” relation. This can be seen in Figure \ref{fig:index-flux-421} where the spectral index (left panel)  and the energy cut-off (right panel) are plotted against the normalization flux for different intrinsic values reported in the literature. The fitted values in this work for Mrk 421 lie between the values from short- and long-term observations. The observed spectra of Mrk 421 reported in the literature that best coincide with our flux points are those reported by VERITAS for a very low activity state \citep{Mrk421VERITAS}, the spectrum measured by MAGIC during the \textit{Fermi} multi-wavelength campaign which was fitted to a single log-parabola \citep{Fermi-421-SED}, and the average spectrum reported by the ARGO-YBJ experiment \citep{Bartoli2016} fitted with a single PL (see Figure \ref{fig:421-all}).

\subsection{Mrk 501}\label{subsec:501-comp}

\begin{figure}[ht!]
\plotone{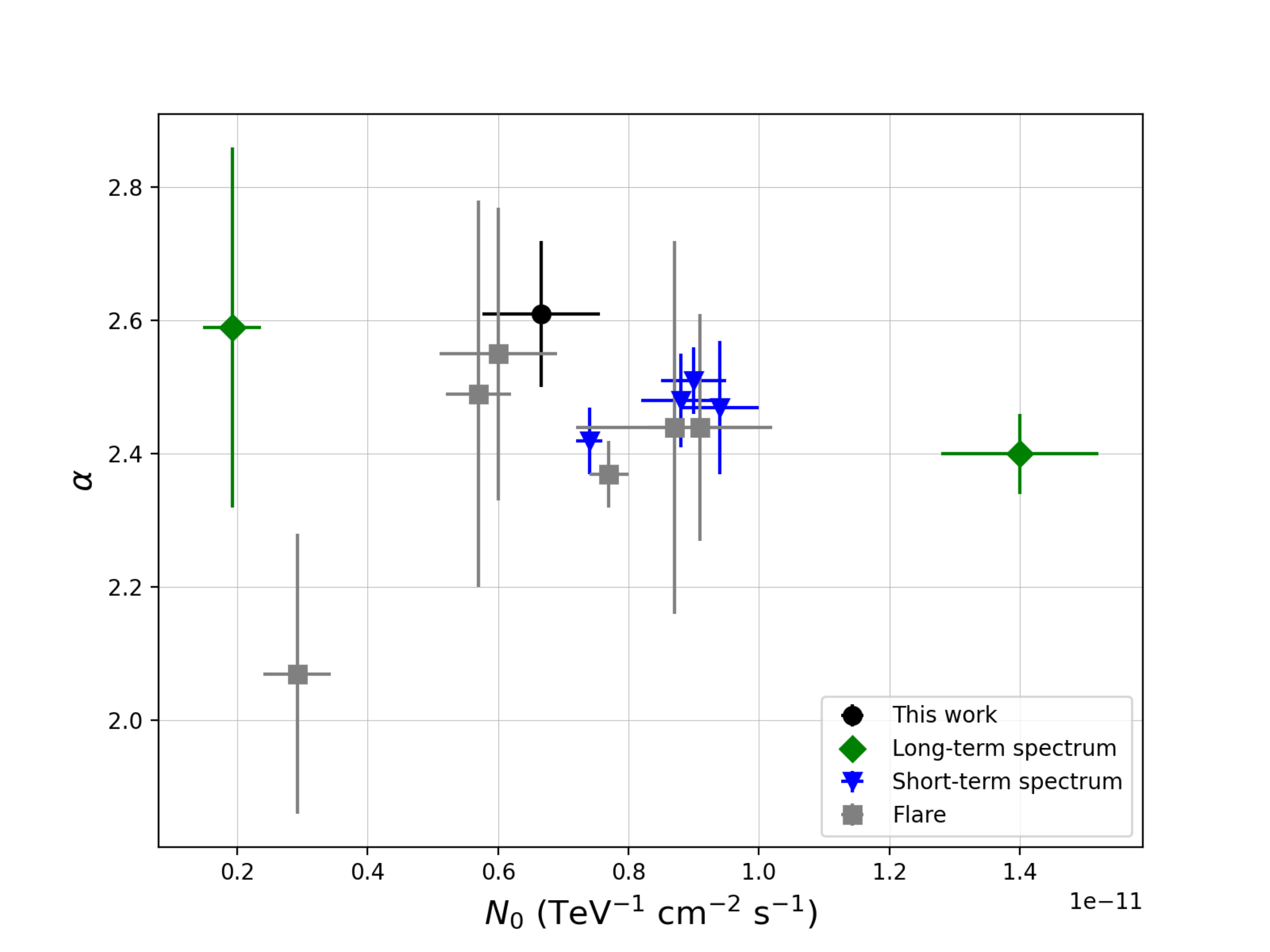}
\caption{Mrk 501. Spectral index vs. normalization flux for reported values in the literature. The black circle correspond to the results in this work, the green diamonds correspond to long-term spectra $> 1$ month) \citep{Bartoli2012, hawc-mrks3}, the blue downward triangles are results reported for short-term ($< 1$ month) \citep{SSC-501-multi2-HS, Fermi-501-SED} spectra, and the grey squares are the reported values from observations when the source presented a high activity state \cite{Bartoli2012, SSC-501-multi2-HS}.\label{fig:index-flux-501}}
\end{figure}

For Mrk 501 the intrinsic spectrum is in agreement with previous results obtained by HAWC \citep{hawc-mrks3}; it is also in agreement with the intrinsic spectrum of the ARGO-YBJ experiment for a 1179.6-day observation period where $\alpha=2.59\pm0.27$ \citep{Bartoli2012}. The trend of having a harder spectrum when the source is in a high activity state is not as noticeable as with Mrk 421, as shown in Figure \ref{fig:index-flux-501} where the spectral index is plotted vs. the normalization flux from reported intrinsic spectra in the literature.

\begin{figure}
\plotone{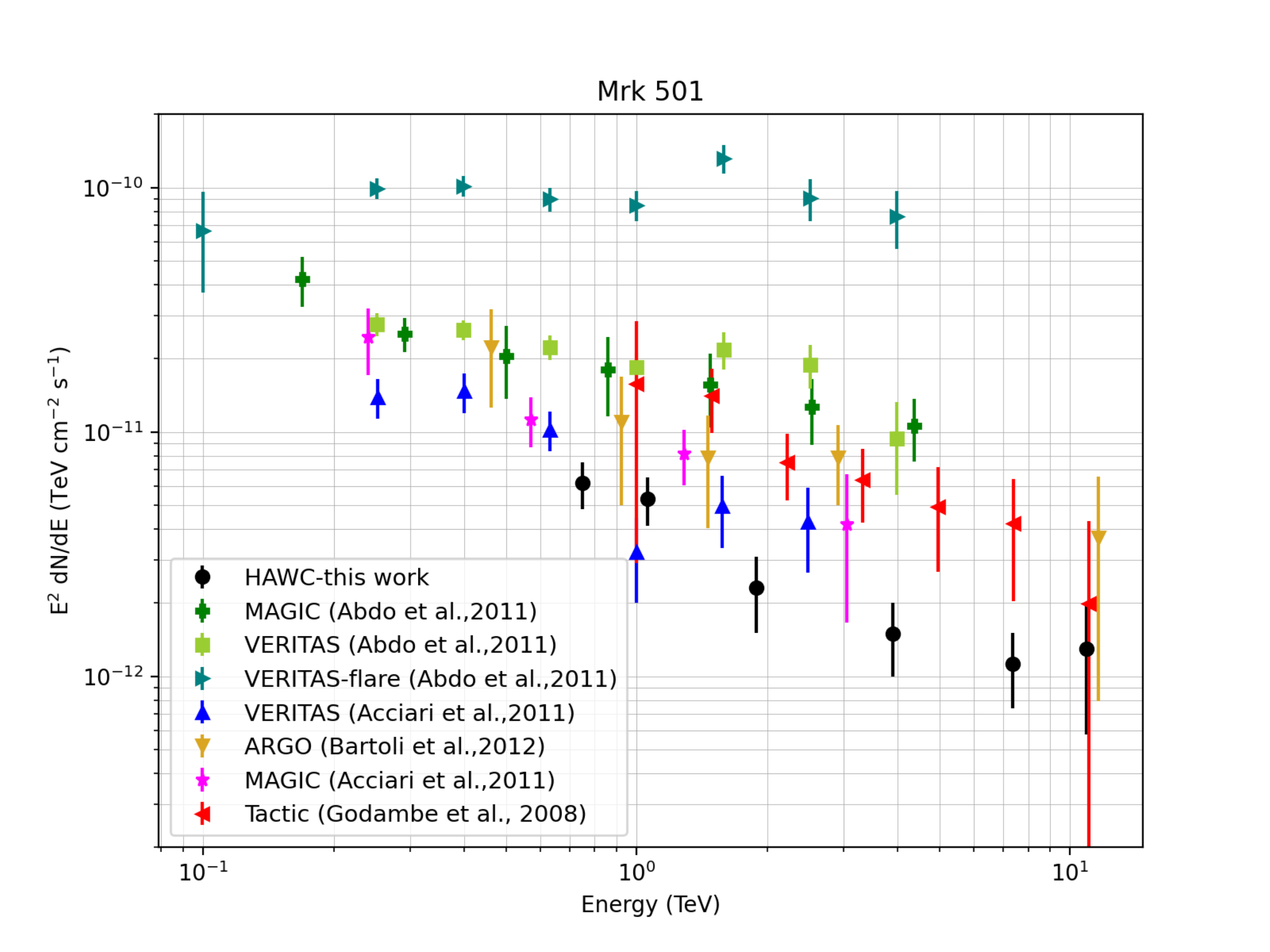}
\caption{VHE spectrum of Mrk 501. The black circles are the HAWC (1038 days) data presented in this work. From the multi-wavelength campaign performed in 2009 \citep{Fermi-501-SED} the green crosses are from MAGIC (16.2 h) and the light green squares are from VERITAS (9.7 h). Also, from this campaign a high activity state was reported by VERITAS (2.4 h) and shown with turquoise right triangles. The blue up triangles and the magenta stars correspond to the spectrum of VERITAS (2.6 h) and MAGIC (1.3 h), respectively, reported in \cite{Mrk501MAGICSED}.The yellow down triangles correspond to the spectrum reported by the ARGO-YBJ experiment (1179.6 days) reported in \cite{Bartoli2012}. And the red left triangles are the data from the TACTIC telescope (129.25 h) reported in \cite{Mrk501TACTIC}. \label{fig:501-all}}
\end{figure}

The energy of the last flux point bin is at 10.90 TeV, which is also one of the highest energy detections for time-averaged spectra to date. The obtained flux points with HAWC data compared to previous observations made with IACTs and ARGO-YBJ, are shown in Figure \ref{fig:501-all}, where the HAWC flux points at higher energies are below previous observations by a factor of $\sim6-7$. This difference can be explained in terms of the Mrk 501 activity state during those observations, such as those reported in \cite{Fermi-501-SED} where a high energy state was detected.

We note that intrinsic spectral parameters depend on the choice of EBL model. For the redshift of Mrk 421 and Mrk 501 ($z = 0.031$ and $z = 0.034$, respectively), and in the detection energy range ($0.5 <E <10$ TeV), the opacity values of the different EBL models that were used in the works cited in sections \ref{subsec:421-comp} and \ref{subsec:501-comp} are not significantly different.
\section{Spectral energy distribution modeling}\label{sec5}
Mrk 421 and Mrk 501 are blazars classified as high-synchrotron–peaked (HBL) BL Lacs  \citep{hbl1, hbl2}. These types of objects are characterized by emitting most of their power in the UV and X-ray range. HBL blazars are also characterized by having a low luminosity, so it is thought that the only seed photons that are scattered at very high energies are synchrotron photons, that is, there is no contribution from external photons from the broad line region, torus or the accretion disk \citep{2008MNRAS.386..945T, agn-rev}. For this reason we assume the the SSC scenario for this work. In addition, SSC models have been widely used to describe the long- \citep{Fermi-421-SED, Fermi-501-SED, Bartoli2011, Bartoli2012, Bartoli2016, Nissim2017}, short-time-averaged \citep{Mrk421MAGIC, Mrk501MAGIC, Mrk501MAGIC-lowstate, MAGIC2010-MRK421, Mrk421VERITAS} SEDs and flares \citep{Mrk501MAGICSED, Bartoli2012, Bartoli2016, Nissim2017} and in general, all describe the observational data well. However, it should be noted that this family of models produce parameter degeneration \citep{2017A&A...603A..31A}, so it is important to be able to restrict some of them, as far as possible.

The electron population within the emission zone has a total energy given by

\begin{equation}
 W_e = \int_{E_{min}}^{E_{max}} E_e\frac{dN_e}{dE_e}dE_e,
\end{equation}
where $E_{min}$ and $E_{max}$ are the minimum and maximum electron energies respectively, and ${dN_e}/{dE_e}$ is the particle distribution embedded in a magnetic field $B$. The electron population is accelerated to relativistic velocities by the magnetic field producing synchrotron radiation, which is then used as a seed photon field for the inverse Compton (IC) scattering. 

The total radiative energy output of the jet $L_{jet}=L_e+ L_p+L_B$ is the sum of the radiative output carried by electrons, protons (under the assumption of one cold proton per emitting relativistic electron) and the magnetic field which are defined as 
\begin{equation}
    L_i \simeq \pi R^2 c\; \Gamma^2 U_i, \quad i=e,p,B
\end{equation}
with $U_e = {3W_e}/{4\pi R^3}$, $U_p=m_pN_p$, and $U_B = {B^2}/{8\pi}$ are the energy densities; being $m_p$ the proton mass \citep{JetPowers}.

\subsection{Multi-frequency data}\label{subsec:sed-data}
\subsubsection{Gamma-ray data}
To construct the VHE part of the SED, we used the HAWC spectra of Mrk 421 and Mrk 501 reported in Sections \ref{subsec:421-res} and \ref{subsec:421-res}.

The high-energy (HE) part of the spectrum, in the MeV-GeV regime, was obtained from the \textit{Fermi} Large Area Telescope (LAT). Using Fermipy \citep{fermipy} we obtained the contemporary spectrum of Mrk 421 and Mrk 501 to cover the IC peak of the SED. The center of the maximum energy bin was set to 0.5 TeV in the configuration file and the contemporary spectrum corresponds to the full-time of the HAWC data set, from June 2015 to July 2018.
\subsubsection{X-ray data}
Contemporaneous data from the X-ray Telescope on board the \textit{Neil Gehrels Swift Observatory} (\textit{Swift}-XRT) \citep{2004ApJ...611.1005G, 2005SSRv..120..165B} is used to complete the X-ray part of the SED of both sources. The data was retrieved using the tools to build \textit{Swift}-XRT data products for point sources via an API \citep{2009MNRAS.397.1177E}) which are then analyzed using the \verb|HEASOFTv.6.29 software|. The selected events were the ones from observations during the windowed timing mode. Using the \verb|XSPEC| package, the 0.3-10 keV average spectrum of each source was fitted to a log-parabola \citep{2004A&A...413..489M} of the form $\frac{dN}{dE}=N_0\times(E/\mbox{keV})^{-(\alpha+\beta\times\log(E/\mbox{keV}))}$ with a fixed hydrogen column density value of $2\times10^{20}\;\mbox{cm}^{-2}$ \citep{2005A&A...440..775K}. 

For Mrk 421, the best fit parameters are $N_0=(2.240\pm0.001)\times 10^{-1}\;\mbox{ph cm}^2\,\mbox{s}^{-1}$, $\alpha=2.070\pm0.005$ and $\beta=0.152\pm0.003$. For Mrk 501, the fitted parameters are $N_0=(3.560\pm0.004)\times 10^{-2}\;\mbox{ph cm}^2\,\mbox{s}^{-1}$, $\alpha=1.865\pm0.002$ and $\beta=0.102\pm0.003$.

\subsubsection{UV, optical and radio data}
For Mrk 421, the \textit{Swift}-UVOT data in the ultraviolet bands \textit{UVW1}, \textit{UVM2} and \textit{UVW2} were taken from \cite{2020ApJS..247...27K} where the observations comprise the non-continuous period between December 2015 and April 2018, making this sample a good contemporary approximation to our data. From optical to radio, the data is taken from \cite{Fermi-421-SED}.

For Mrk 501, data from \textit{Swift}-UVOT in the \textit{UVW2} band were taken from \cite{2021A&A...655A..93A}, covering a non-continuous period between June 2015 and April 2018. From optical to radio the data is taken from \cite{Fermi-501-SED}.

\subsection{SED modeling}
We use \verb|agnpy|\footnote{\url{https://agnpy.readthedocs.io/en/latest/index.html}}, a Python package to calculate the photon spectra produced by leptonic radiative processes in AGN. \verb|agnpy| bases the synchrotron and SSC processes in the work published by \cite{2009herb.book.....D} and \cite{2008ApJ...686..181F}. A ${\chi}^2$ fit to the multi-frequency data is performed using \verb|sherpa|\footnote{\url{https://sherpa.readthedocs.io/en/latest/index.html}}.

The tested electron energy distributions were a single power-law, a broken power-law and a log-parabola, where the spectral parameters, such as the normalization flux, spectral indices, curvature and energy break, and $E_{min}$ and $E_{max}$ were left free to vary. For the synchrotron and IC flux calculation we also left the magnetic field $B$ free in the fit. Since this work does not contemplate any variability studies, the size of the emission zone can not be constrained by this quantity and, therefore, is fixed to a value of $R = 5 \times 10^{16}$ cm for Mrk 421 and $R = 10^{17}$ cm for Mrk 501, based on the works on long-term averaged spectra of these sources in the past \citep{Fermi-421-SED, Fermi-501-SED, Bartoli2011, Nissim2017}. The numerical computation is performed in the co-moving frame so the Doppler factor can be fitted as a free parameter.

\subsection{SED modeling results.}\label{subsec:sed-res}
\subsubsection{Mrk 421}
\begin{figure}[ht!]
\plotone{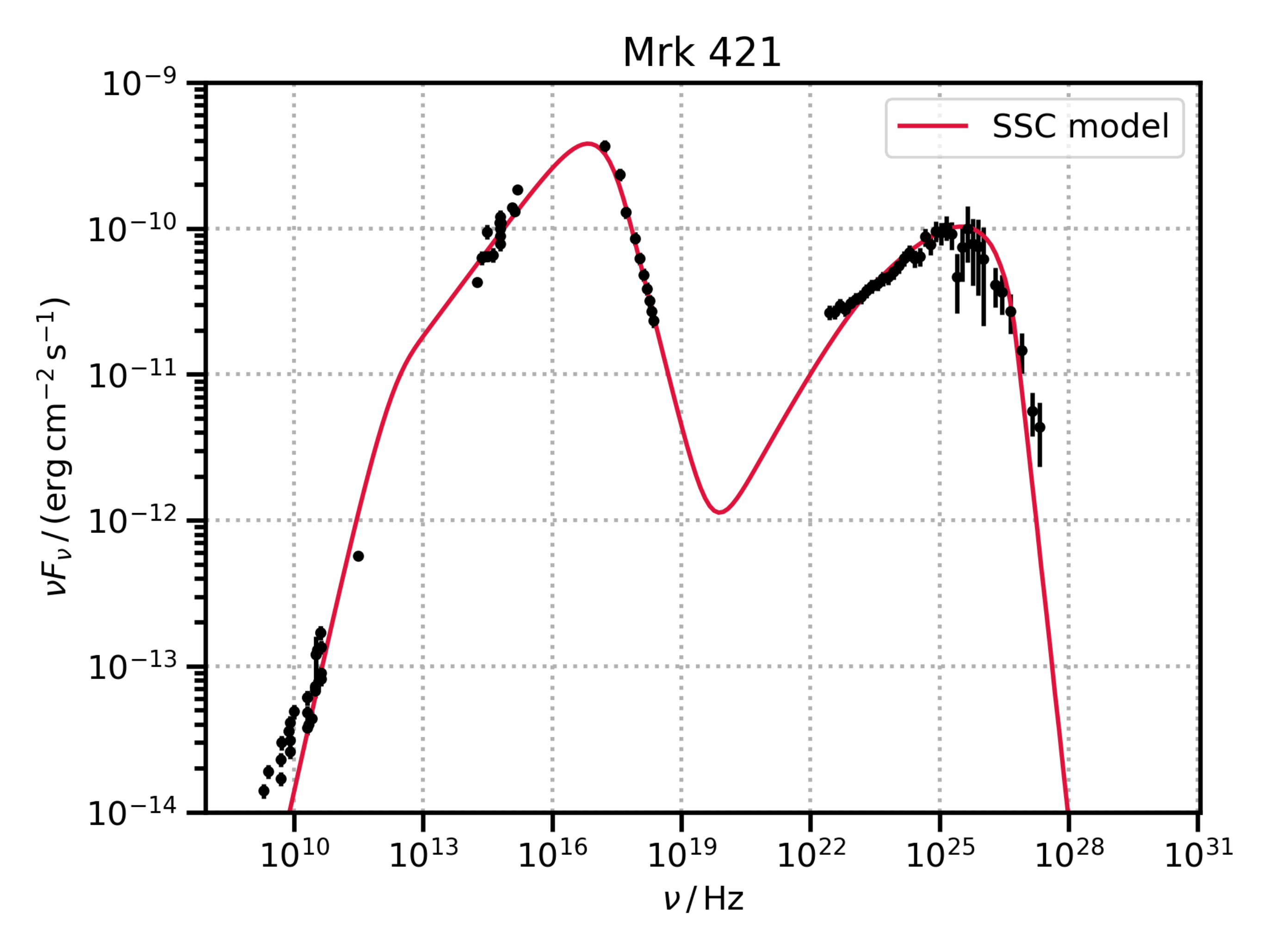}
\caption{Mrk 421 SED. The data points are described in Section \ref{subsec:sed-data}. The best SSC model (red curve) corresponds to a broken power law and the resulting parameters are shown in Table \ref{tab:para-sed-421}.\label{fig:sed-model-421}}
\end{figure}

For Mrk 421, the best fit corresponds to a Doppler factor of $\delta=25\pm1$ for an electron energy distribution that follows a broken power-law with energy break $E_{break} = 112 \pm 41$ GeV and spectral indexes before and after the break of $\alpha_1=2.21\pm0.01$ and $\alpha_2 = 5.4 \pm 0.1$, respectively. The minimum and maximum electron energies are best fit to $E_{min} = 425 \pm 8$ MeV and $E_{max} = 51 \pm 13$ TeV. The magnetic field results in a value of $B = 24 \pm 6$ mG. Figure \ref{fig:sed-model-421} shows the multi-frequency SED of Mrk 421 and the best SSC model (red line). Table \ref{tab:para-sed-421} shows a summary of the SED fitting results.

\begin{table}[ht]
\centering
\caption{Fitted parameters in the SSC leptonic model for Mrk 421. } \label{tab:para-sed-421}
\begin{tabular}{l c c }
\hline\hline
Parameter & Symbol & Mrk 421\\
\hline
Doppler factor & $\delta$ & $25\pm1$\\
Magnetic field & $B$ [mG] & $24\pm6$ \\
Spectral index before break & $\alpha_1$ & $2.21\pm0.01$  \\
Spectral index after break & $\alpha_2$ & $5.4\pm0.1$  \\
Energy break & $E_{break}$ [GeV] & $112\pm41$ \\
Minimum electron energy & $E_{min}$ [MeV] & $425\pm8$ \\
Maximum electron energy &  $E_{max}$ [TeV] & $51\pm13$ \\
Jet power in electrons & $L_e\;\times 10^{44}$ [erg $\mbox{s}^{-1}$] & $2.3$ \\
Jet power in protons & $L_p\;\times 10^{44}$ [erg $\mbox{s}^{-1}$] & 4.0 \\
Jet power in magnetic field & $L_B\;\times 10^{42}$ [erg $\mbox{s}^{-1}$] & $6.8$ \\
\hline \hline
\end{tabular}
\end{table}

\begin{table}[ht]
\centering
\caption{Comparison between previous SSC parameters, $\delta$, $B$ and $R$, and results in this work for Mrk 421.} \label{tab:comp-sed-421}
\begin{tabular}{c c c c c}
\hline\hline
$\delta$ & $B$ & $R$ & Flux state & Reference\\
& [mG] & $\times 10^{16}\;\mbox{[cm]}$ & & \\
\hline
15 & 200 & 1.1 &  Low & \cite{Mrk421MAGIC}\\
15 & 150 & 5 & High & \cite{Bartoli2011}\\
16 & 80 &  5 & Low &  \\
40 & 200 & 0.25 & Low-Mid-High & \cite{Mrk421VERITAS}\\
21 & 38 & 5.2 & Long-term averaged & \cite{Fermi-421-SED}\\
$38_{-4}^{+6}$ & $48\pm0.012$ & 1 & Long-term averaged & \cite{Bartoli2016}\\
$25\pm1$ & $24\pm6$ & $5$ & Long-term averaged & This work\\
\hline \hline
\end{tabular}
\end{table}

\subsubsection{Mrk 501}
\begin{figure}[ht!]
\plotone{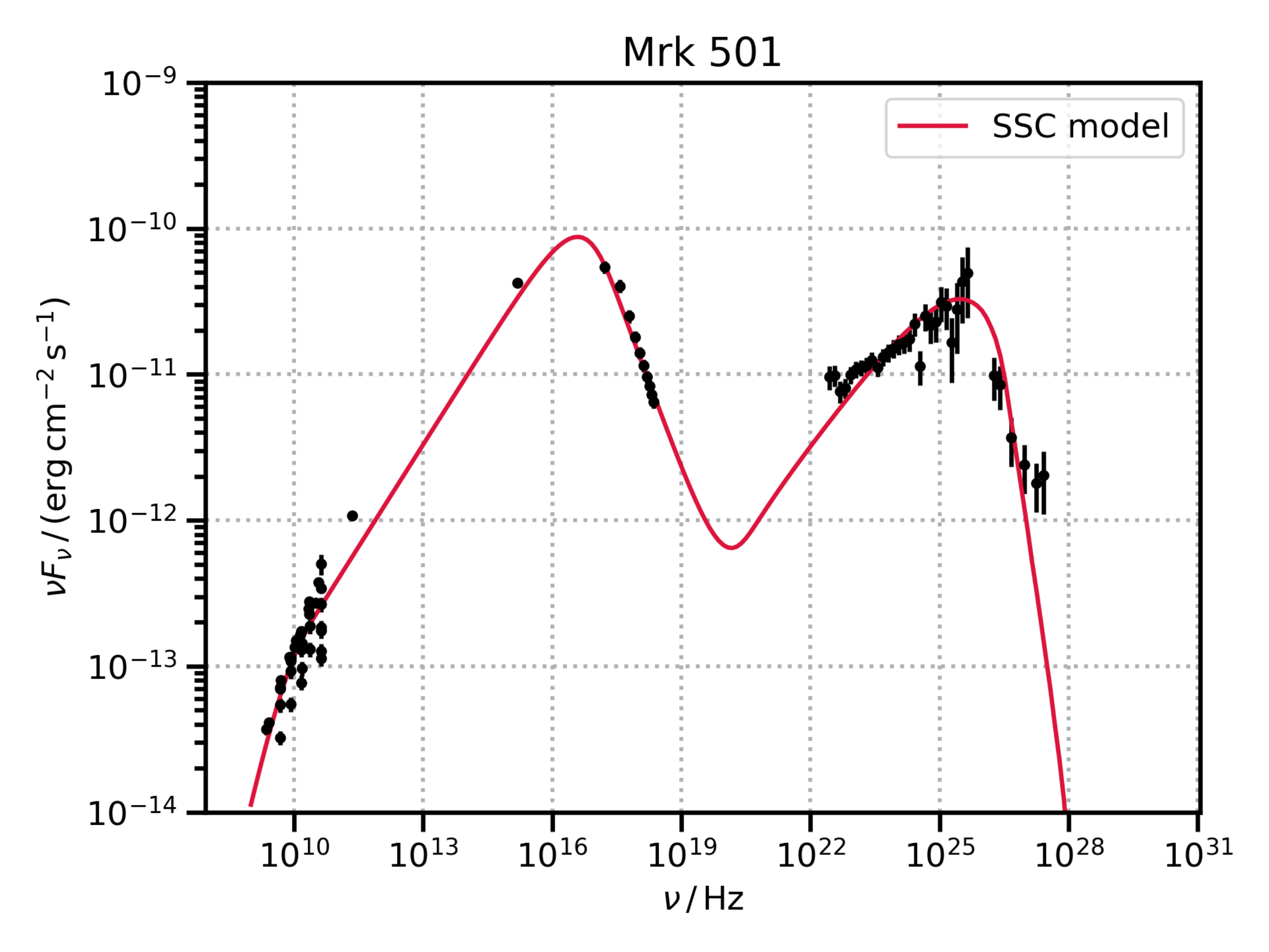}
\caption{Mrk 501 SED. The data points are described in Section \ref{subsec:sed-data}. The best SSC model (red curve) corresponds to a broken power law and the resulting parameters are shown in Table \ref{tab:para-sed-501}.\label{fig:sed-model-501}}
\end{figure}

For Mrk 501 the SED is better described with Doppler factor value of $\delta = 13\pm0.7$. The electron energy distribution that results in a better fit is a broken power-law with energy break $E_{break} = 470\pm20$ GeV and spectral indexes before and after the break of $\alpha_1 = 2.1 \pm 0.01$ and $\alpha_2 = 4.6 \pm 0.1$, respectively. The best fit values of the minimum and maximum electron energy are $E_{min} = 166\pm8$ MeV and $E_{max} = 19\pm0.5$ TeV. According to the IC scattering process, the energy of the VHE photons must not exceed that of the electrons, so accounting for the Doppler boosting, the $E_{max}$ value agrees with our observations. The magnetic field is fitted to a value of $B = 20 \pm 5$ mG. This model is shown in Figure \ref{fig:sed-model-501} (red line) and a summary of these results in Table \ref{tab:para-sed-501}.

\begin{table}[ht]
\centering
\caption{Fitted parameters in the SSC leptonic model for Mrk 501. } \label{tab:para-sed-501}
\begin{tabular}{l c c}
\hline\hline
Parameter & Symbol & Mrk 501\\
\hline
Doppler factor & $\delta$ & $13\pm0.7$\\
Magnetic field & $B$ [mG] & $20\pm5$\\
Spectral index before break& $\alpha_1$ & $2.1\pm0.01$ \\
Spectral after before break& $\alpha_2$ & $4.6\pm0.1$ \\
Energy break & $E_{break}$ [GeV] & $470\pm20$ \\
Minimum electron energy & $E_{min}$ [MeV] & $166\pm8$\\
Maximum electron energy &  $E_{max}$ [TeV] & $19\pm0.5$ \\
Jet power in electrons & $L_e\;\times 10^{44}$ [erg $\mbox{s}^{-1}$] & $2.7$\\
Jet power in protons & $L_p\;\times 10^{44}$ [erg $\mbox{s}^{-1}$] & 4.1\\
Jet power in magnetic field & $L_B\;\times 10^{42}$ [erg $\mbox{s}^{-1}$] & $4.1$\\
\hline\hline
\end{tabular}
\end{table}

\begin{table}[ht]
\centering
\caption{Comparison between previous SSC parameters, $\delta$, $B$ and $R$, and results in this work for Mrk 501.} \label{tab:comp-sed-501}
\begin{tabular}{c c c c c}
\hline\hline
$\delta$ & $B$ & $R$ & Flux state & Reference\\
& [mG] & $\times 10^{16}\;\mbox{[cm]}$ & & \\
\hline
25 & 310 & 0.1 &  Low & \cite{Mrk501MAGIC}\\
20 & 313 & 0.103 & Low & \cite{Mrk501MAGIC-lowstate}\\
12 & 15 & 13 & Long-term averaged & \cite{Fermi-501-SED} \\
12 & 70 & 3 & Long-term averaged & \cite{Bartoli2012}\\
10 & 100 & 3 & High & \\
$13\pm0.7$ & $20\pm5$ & $10$ & Long-term averaged & This work\\
\hline \hline
\end{tabular}
\end{table}

In Tables \ref{tab:comp-sed-421} and \ref{tab:comp-sed-501} we provide a comparison between our results and previous analysis using a SSC model for Mrk 421 and Mrk 501, respectively. As can be seen, the most notable difference lies in the value of the magnetic field, which is up to an order of magnitude larger than our results for analyses that were carried out using VHE data averaged over short periods or flares. This can also be noted for the size of the emission zone of Mrk 501, whose value is up to two orders of magnitude larger for long-term averaged spectra. The small difference between our results and those in the literature that include VHE data averaged over long periods of time, show a good agreement between the outcome of the newly analysed HAWC data presented here and those previously reported from other experiments.

The jet power in electrons for both sources is comparable to that of protons, $L_e\sim L_p$, and both are larger than the jet power carried by the magnetic field, $L_e>L_B$, thus the Poynting flux does not contribute significantly to the total radiation of the jet. The total radiative energy output of Mrk 421 jet is then $L_{jet-421}=6.5\times 10
^{44}$ erg $\mbox{s}^{-1}$ which corresponds to $\sim4\%$ of the Eddington luminosity and for Mrk 501, $L_{jet}=6.1\times 10^{44}$ erg $\mbox{s}^{-1}$, which represents the $\sim0.3\%$ of its Eddington luminosity.
\section{Summary and outlook}\label{sec6}
We report the detection of Mrk 421 and Mrk 501, above 0.5 TeV with the High Altitude Water Cherenkov (HAWC) Gamma-Ray Observatory using 1038 days of exposure comprising the period between June 2015 and July 2018. 

\begin{enumerate}

    \item For Mrk 421, the VHE intrinsic spectrum is well described by a power law with an exponential energy cut-off. For Mrk 501 the intrinsic VHE spectrum is described by a single power law. 
    \item These results are in good agreement with those previously obtained with HAWC on both sources once the EBL attenuation is taken into account. Additionally, the reported values for the intrinsic spectra in this work are compatible with those in previous averaged spectra reported by IACTs and EAS experiments, setting this way a baseline energy spectrum of each source. It is also important to mention that the obtained flux points in this work are in good agreement with the observed spectra reported in the literature for Mrk 421; however, for Mrk 501 the HAWC flux points lay below the observed spectra reported in the literature, this could be related to the activity state of the source when it was observed in the past.
    \item Compared to previously published results using HAWC data, this is the first time that we estimate the highest energy of the detected signal, with 9 TeV for Mrk 421 and 12 TeV for Mrk 501 at a $2\sigma$ confidence level, which for both sources, is one of the highest energy detections reported to date, for spectra averaged over long periods of time. This contributes to the restriction of the energy detection limits of both sources.
    \item The SED built using contemporaneous data from HAWC, \textit{Fermi}-LAT, \textit{Swift}-XRT and \textit{Swift}-UVOT, along with previously published data in the radio to optical energy range, was modeled using a one-zone SSC scenario.
    \item The estimated physical parameters from the jet are in general agreement with values found in the literature for long-term observations, confirming that both sources are intrinsically different assuming that the same physical processes take place.
\end{enumerate}
  
To characterize the spectrum at VHE of Mrk 421 and Mrk 501 in greater detail, it is important to identify the periods of variability of both sources and thus carry out the spectral analysis in each of them, this way the physical processes that give rise to these energy flux variations can be constrained. To achieve this, a time resolved analysis of the data set used in this work is necessary and will be addressed in future publications.
\section*{Acknowledgment}
We acknowledge the support from: the US National Science Foundation (NSF); the US Department of Energy Office of High-Energy Physics; the Laboratory Directed Research and Development (LDRD) program of Los Alamos National Laboratory; Consejo Nacional de Ciencia y Tecnolog\'ia (CONACyT), M\'exico, grants 271051, 232656, 260378, 179588, 254964, 258865, 243290, 132197, A1-S-46288, A1-S-22784, c\'atedras 873, 1563, 341, 323, Red HAWC, M\'exico; DGAPA-UNAM grants IG101320, IN111315, IN111716-3, IN111419, IA102019, IN112218; VIEP-BUAP; PIFI 2012, 2013, PROFOCIE 2014, 2015; the University of Wisconsin Alumni Research Foundation; the Institute of Geophysics, Planetary Physics, and Signatures at Los Alamos National Laboratory; Polish Science Centre grant, DEC-2017/27/B/ST9/02272; Coordinaci\'on de la Investigaci\'on Cient\'ifica de la Universidad Michoacana; Royal Society - Newton Advanced Fellowship 180385; Generalitat Valenciana, grant CIDEGENT/2018/034; Chulalongkorn University’s CUniverse (CUAASC) grant. Thanks to Scott Delay, Luciano D\'iaz and Eduardo Murrieta for technical support. This work made use of data supplied by the UK Swift Science Data Centre at the University of Leicester.

\bibliography{main}{}
\bibliographystyle{aasjournal}

\end{document}